\documentclass[conference]{IEEEtran}
\usepackage{cite}
\usepackage{amsmath,amssymb,amsfonts}
\usepackage{graphicx}
\usepackage{textcomp}
\usepackage{xcolor}
\usepackage{multirow}
\usepackage{multicol}
\usepackage{graphicx}
\usepackage{booktabs}
\usepackage{xcolor}
\usepackage{enumitem}
\usepackage[utf8]{inputenc}
\usepackage{tabularx} 
\usepackage{threeparttable}

\def\BibTeX{{\rm B\kern-.05em{\sc i\kern-.025em b}\kern-.08em
    T\kern-.1667em\lower.7ex\hbox{E}\kern-.125emX}}

\begin{document}
\bstctlcite{IEEEexample:BSTcontrol}
\title{Audio-Visual Turn-taking Prediction \\ in Cocktail Party Scenarios
\thanks{Funding agency information will be disclosed upon acceptance.}
}


\author{\IEEEauthorblockN{Long-Vu Hoang \hspace{2em} Naomi Harte}
\IEEEauthorblockA{\textit{Sigmedia Group, School of Engineering, Trinity College Dublin, Ireland} \\
\{lhoang, nharte\}@tcd.ie}
}

\maketitle

\begin{abstract}

Current predictive turn-taking models (PTTMs) achieve strong performance on benchmarks with controlled acoustic conditions and clean audio signals. Their generalisation to conversations with overlapping speech and background interference remains underexplored. In this research, we evaluate audio-visual PTTMs trained with clean data on a challenging cocktail-party testbed derived from the AVCocktail dataset, and analyse their adaptation behaviour to this new domain. Experimental results show consistent performance degradation across audio and visual modalities under noisy conditions, with up to 38\% relative drop in weighted F1. Fine-tuning on the new domain improves robustness, but gains vary across modalities and depend on the size of the available pre-training data. These findings provide insights into the different generalisation and adaptation capabilities of the audio and visual modalities, and indicate the need for robust modelling strategies to adapt to the complexities of human interactions in noise. All code and turn labels are made publicly available to facilitate further research\footnote{Code and labels at: https://github.com/lggvu/mm-turn-taking.}.

\end{abstract}


\section{Introduction}
Turn-taking is a systematic organisation of who speaks when in a conversation. In human-human dialogue, there is an average silence of $200$ms between turns, while language production takes around $600$ms \cite{levinson2015timing}. Therefore, speakers need to anticipate when to take the floor to maintain a natural conversation flow, making turn-taking a predictive problem. Modelling such behaviour remains challenging for machines \cite{skantze_turn-taking_2021, castillo-lopez_survey_2025}. Early approaches relied on silence-thresholding heuristics, which often fail in natural, realistic conversations \cite{skantze_turn-taking_2021}. Recently, predictive turn-taking models (PTTMs) have emerged as an effective solution. These models leverage the strong ability to learn rich representations from continuous signals by sequence-to-sequence architectures,
with the Transformer \cite{vaswani2017attention} being the standard backbone
\cite{ekstedt22_interspeech, inoue2024multilingual, russell2025visual, onishi_multimodal_2023}.

PTTMs have demonstrated strong performance across a range of conversational benchmarks in quiet, controlled environments \cite{castillo-lopez_survey_2025}. These include dyadic dialogue over telephone like Switchboard \cite{godfrey1992switchboard}, or structured interactions in controlled environments like NoXi \cite{cafaro2017noxi} and Spontaneous Speech Corpus of Japanese \cite{maekawa2000spontaneous}. Dyadic videoconferencing datasets such as Candor \cite{reece2023candor} and Japanese Travel Agency Task \cite{inaba2022collection} have also been used to evaluate turn-taking prediction models. Recent work has begun to explore the performance of PTTMs under noisy real-world conditions. Inoue et al. \cite{inoue2025noise} conducted a field experiment in a noisy shopping mall to measure the performance of PTTMs using subjective and objective metrics, and augmented the training data with synthetic background noise to improve robustness. Russell and Harte \cite{oconnorrussell25_interspeech} investigated the effects of different types of noise in both training and evaluation, and showed that the addition of visual cues and synthetic noise to training enables more robust turn-taking in noise.
However, no study has analysed how PTTMs generalise and adapt to noisy environments and more complex human interactions, particularly under cocktail-party conversations, where overlapping speech and background noise are the norm. 

The cocktail-party effect refers to the ability to focus one's listening attention on one conversation while ignoring others, under the effect of background noise. Cocktail-party scenarios challenge PTTMs not only in terms of noisy acoustic signals, but also in differences in speech production and visual expressions of speakers compared to quiet environments \cite{garnier2018hyper, vsimko2016hyperarticulation}. Various studies have discussed how noise exposure triggers the Lombard effect - an adaptation in speech articulation, with e.g. increased vocal intensity and fundamental frequency \cite{garnier2014speaking}. Speakers may also adapt their visual aspects of speech, such as lip movements, leading to different distributions of visual cues \cite{trujillo2021speakers, vsimko2016hyperarticulation}. Therefore, the cocktail-party phenomenon poses significant domain differences in both audio and visual modalities compared to conversations in quiet, controlled environments, or augmented with synthetic noise that speakers do not experience during recording.

In this study, we aim to answer two research questions:
\textit{RQ1: Can PTTMs trained on clean, controlled datasets generalise to cocktail-party scenarios, and do audio and visual modalities exhibit similar generalisation behaviour?} To address this, we extensively benchmark audio-visual models trained with the clean Candor data \cite{reece2023candor} on a challenging cocktail-party dataset from speech recognition literature, AVCocktail \cite{nguyen25b_interspeech}. We proposed an automated procedure to deduce reliable turn-taking labels from the transcriptions in AVCocktail \cite{nguyen25b_interspeech}. 
\textit{RQ2: We investigate whether a simple fine-tuning can adapt models effectively to the cocktail-party scenario in both audio and visual modalities.} From these initial results, we analyse how each modality adapts to the new domain, and ablate the effect of pre-training dataset size on the adaptation.

The remaining sections of this paper are organised as follows. Section \ref{sec:problem-def} provides a formal introduction to the turn-taking problem and the datasets used for our experiments. Section \ref{sec:exps} details the models and the setups for training and evaluation. The results are discussed in Section \ref{sec:results}. We summarise and conclude our research in Section \ref{sec:conclusion}.

\section{The Turn-Taking Prediction Problem}
\label{sec:problem-def}
\subsection{Problem Formulation}

There is no single definition of a \textit{turn}. Conversation analysts define a turn as a social action, such as a question or an agreement \cite{sacks1974simplest}. In speech technology, a turn is defined as a sequence of inter-pausal units (IPUs) from a speaker, which are not interrupted by IPUs from any other speaker. IPUs are stretches of audio from one speaker without any silence exceeding a certain amount, typically $200$ms \cite{skantze_turn-taking_2021} or $250$ms \cite{russell2025visual}.
After each IPU, the speaker holding the floor can either yield the turn to another speaker (\textit{shift}), or continue the turn (\textit{hold}).

\subsection{The Candor Dataset}
To train and evaluate PTTMs models, we use the audio-visual Candor \cite{reece2023candor} dataset. Candor has 850 hours of 1,657 unscripted dyadic videoconferencing conversations in English, recorded in 30-FPS videos and 16kHz stereo audio. Following the event definitions from \cite{oconnorrussell25_interspeech}, silences of over $250$ms are first selected, then shifts and holds are defined by considering if the speakers before and after the silence are the same or not.  We use the training and development subset of the first fold in \cite{russell2025visual, oconnorrussell25_interspeech}, whose statistics are listed in Table \ref{tab:datasets-stats}.

\subsection{The AVCocktail Dataset}
\label{sec:avcocktail-dataset}
\begin{figure}[t]
    \centering
    \begin{minipage}{0.38\linewidth}
        \centering
        \includegraphics[width=\linewidth]{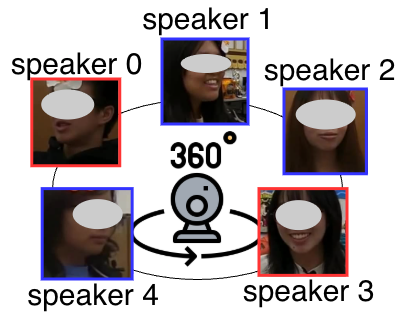}
        \\
    \end{minipage}
    \hfill
    \begin{minipage}{0.42\linewidth}
        \centering
        \includegraphics[width=\linewidth]{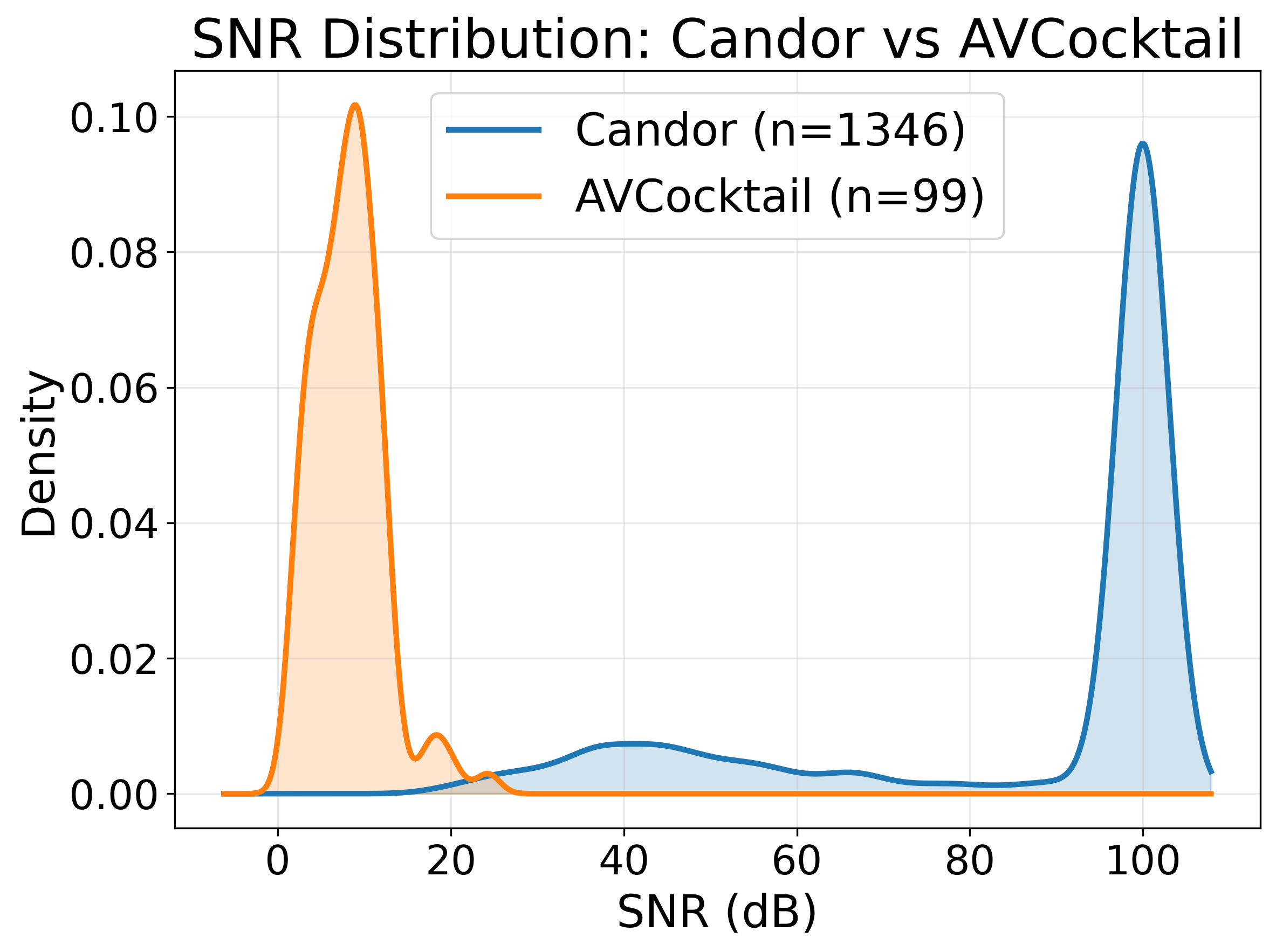}
        \\
    \end{minipage}
    \caption{\textbf{Left:} An example schematic of a session in AVCocktail with two parallel conversations and 5 speakers. Faces are blurred due to licence reasons. Speakers in the same border colour engage in the same dialogue. Only dyads are chosen for experiments (red in this example), \texttt{speaker 0} and \texttt{speaker 3} are assumed to be Left and Right speakers, respectively. \textbf{Right:} The estimated SNR of Candor and AVCocktail (train and dev, totalling $1,346$ recordings for Candor and $99$ for AVCocktail).}
    \vspace{-1em}
    \label{fig:side-by-side}
\end{figure}
\begin{table}[htbp]
\vspace{-1em}
\small
\centering
\caption{Statistics of the Candor and AVCocktail datasets.}
\label{tab:datasets-stats}
\resizebox{\columnwidth}{!}{%
\begin{tabular}{@{}lrrrrrr@{}}
\toprule
\textbf{Subset} &
  \textbf{\# Hours} &
  \textbf{\begin{tabular}[c]{@{}r@{}}\# Dyads\end{tabular}} &
  \textbf{\# Shifts} &
  \textbf{\# Holds} &
  \textbf{\begin{tabular}[c]{@{}r@{}}Avg \# Shifts\\ /min\end{tabular}} &
  \textbf{\begin{tabular}[c]{@{}r@{}}Avg \# Holds\\ /min\end{tabular}} \\ \midrule
Candor-train     & 601.81     & 1,077 & 52,721 & 130,758 & 1.47  & 3.63 \\
Candor-dev       & 153.42     & 269   & 14,062 & 34,592  & 1.54  & 3.77 \\ \midrule
AVCocktail-train & 6.15 & 62    & 9,462  & 5,918   & 13.37 & 8.68 \\
AVCocktail-dev   & 3.55 & 35    & 2,116  & 1,757   & 9.93  & 8.25 \\ \bottomrule
\end{tabular}%
}
\vspace{-1em}
\end{table}

AVCocktail \cite{nguyen25b_interspeech} is a dataset of multi-party conversations in cocktail-party scenarios. Each session involves speakers having conversations around a table in the same room (Figure \ref{fig:side-by-side}). The room layout and speakers involved vary across the sessions. Recordings are approximately $5-7$ minutes long and can involve a maximum of $8$ active speakers and $4$ parallel conversations, pushing the speech overlap ratio up to $100\%$ at some moments. A 360-degree camera centred at the discussion table produces videos cropped to 
$224\times224$ pixels surrounding each speaker's face and single-channel audio signals of all speakers at $16$kHz. Each speaker also wears a lapel microphone, providing clearer audio only for ASR and timestamp annotations, but not for training and evaluation.

\begin{figure}
    \centering
    \includegraphics[width=0.75\linewidth]{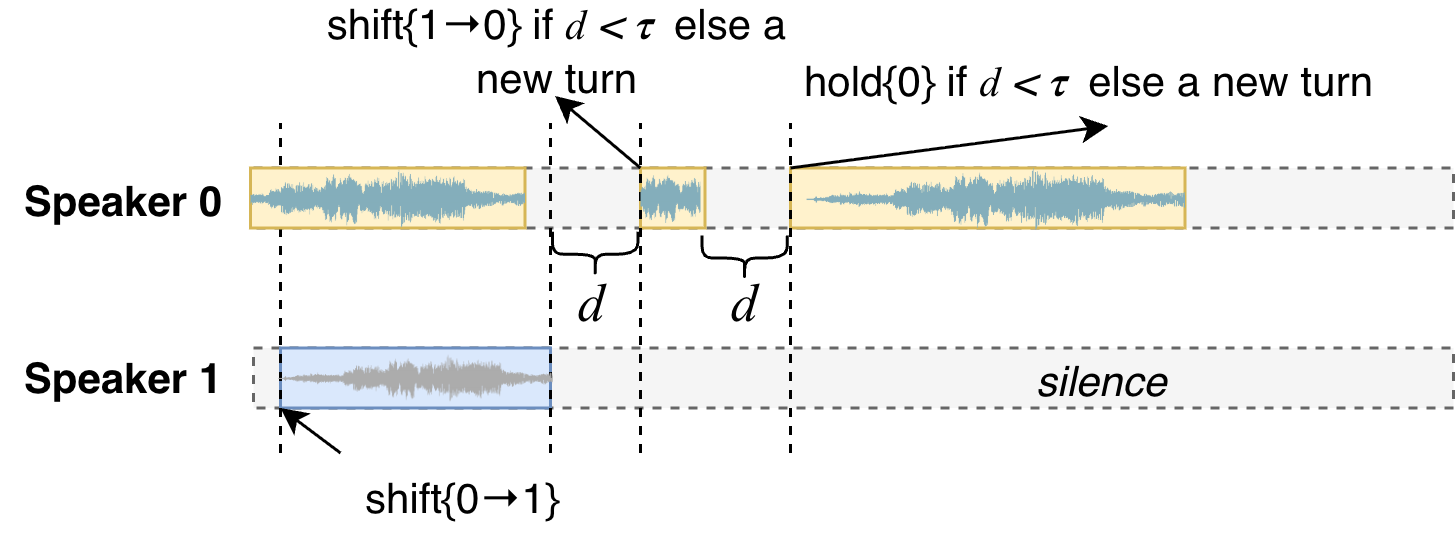}
    \caption{The definitions of shift and holds in the AVCocktail dataset. Grey chunks are silence regions. Only sentences longer than 1 second are considered. The starting timestamps of sentences are candidates for the time of a shift or hold event. Speech regions further than $\tau$ seconds away are considered new turns. }
    \label{fig:turn-def}
    \vspace{-1em}
\end{figure}

AVCocktail was originally developed for audio-visual speech recognition (AVSR). We adapt the dataset to turn-taking prediction since it represents a more realistic and challenging conversational setting compared to current literature. In a cocktail party, overlapping speech, background noise, frequent speaker interruptions, and competing visual cues make it considerably harder to identify turn boundaries and speaker intent, compared to the clean recording conditions like Candor. The dataset is also provided with manually annotated sentence-level transcriptions, facilitating the automatic derivation of shift/hold labels. To the best of our knowledge, it is the only dataset with sufficient features for an automatic evaluation of audio-visual turn-taking prediction in a cocktail-party scenario.

The processing steps to adapt AVCocktail to a turn-taking problem are detailed as follows. 
Firstly, dyads are chosen from the pool of parallel conversations in a session. The corresponding video crops and single-channel audio are extracted.
Secondly, the single-channel audio is duplicated into a stereo audio. The stereo audio and the cropped face streams are then used as inputs to the turn-taking prediction models.

The most important step is to establish reliable turn labels automatically from AVCocktail. A turn is constructed from sequences of IPUs \cite{skantze_turn-taking_2021}, which can be effectively defined from fine-grained word-level transcriptions \cite{russell2025visual}. However, due to the high overlapping speech ratio in AVCocktail, our attempts to obtain reliable word-level alignments using forced alignment were unsuccessful. The event definition is carried out following the convention in Figure \ref{fig:turn-def}. First, sentences of at least $1$ second long are chosen to avoid considering backchannels. Each starting timestamp of a sentence will be considered a candidate for the moment of a shift or hold event. 
If a candidate timestamp is contained in the speaking time range of another speaker, it is considered a \textit{shift}.
Otherwise, it is a \textit{shift} if less than $\tau$ seconds from the ending speaking time of another speaker; or a \textit{hold} if not over $\tau$ seconds from the ending time of the same speaker. Any sentences more than $\tau$ seconds away are considered a new turn.
Prior research typically sets the threshold $\tau$ to $200$–$250$ ms, reflecting the average silences between human speaking turns \cite{skantze_turn-taking_2021}. In contrast, we use a more relaxed threshold of $2.0$s for AVCocktail, based on manual inspection. This choice mitigates both missed true events and excessive false positives, as AVCocktail annotations are at the sentence level and their starting and ending timestamps can deviate from actual speech boundaries by $0.5$–$1$s.

\subsection{Domain Shift Between Candor and AVCocktail}

From the dataset statistics in Table \ref{tab:datasets-stats},  AVCocktail exhibits substantially more complex interaction dynamics than Candor. The average shifts per minute in AVCocktail-train is approximately $1.5$ times higher than the number of holds, contrasting with the distribution in Candor-train. This suggests that speakers in cocktail-party environments engage in more rapid and dynamic turn exchanges. 
The distributions of signal-to-noise ratio (SNR) estimated with WADA-SNR  \cite{kim2008robust} are visualised in Figure \ref{fig:side-by-side}. The median SNR in Candor is in the range of very clean signals, $90-100$dB, with a secondary mode at $40$dB. AVCocktail consistently exhibits noisier audio signals with median SNR values centred around $15$ dB, which can pose considerably more challenges to PTTMs.
\begin{figure}
    \centering
    \includegraphics[width=\linewidth]{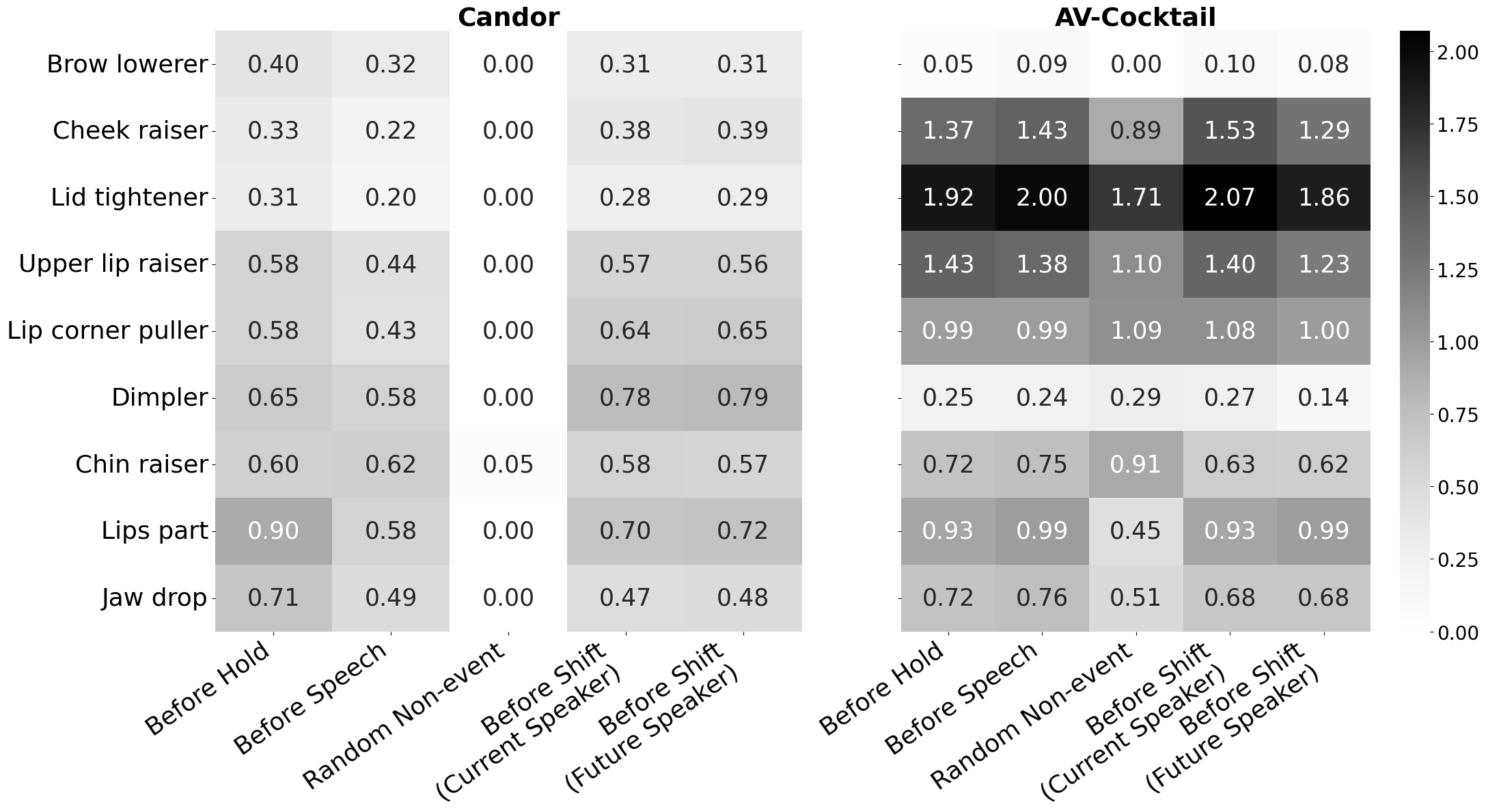}
    \caption{Heatmaps of median intensity value (number in the cells) for each FAU (vertical axis) in Candor-train and AVCocktail-train.}
\vspace{-1em}
    \label{fig:heatmaps-of}
\end{figure}
Another key difference between Candor and AVCocktail is the interaction setting: Candor is videoconferencing (VC), where participants see each other only through screens, whereas AVCocktail involves in-person interaction. This may result in major shifts in the visual information presented by speakers. VC increases cognitive load and reduces gaze exchange over time \cite{han2024person}. People are also more facially expressive in person than in VC, reflected in the higher intensity of facial action units (FAUs) \cite{rollings2024facial}. FAUs \cite{ekman1978facial} are a set of facial muscle movements (jaw drops, eyelid tightening, eyebrow raising, etc.) that represent complex facial expressions.

To analyse the differences in visual cues between the two datasets, we compare the FAUs intensity extracted by OpenFace \cite{baltruvsaitis2016openface}. Following the analysis convention in \cite{russell2025visual}, we select a set of representative FAUs with the highest intensity before and after shift and hold events, (see the y-axis in Figure \ref{fig:heatmaps-of}), and report their median intensity values over a $200$ms window.
We additionally sample an equal number of regions from random non-event intervals, ensuring these are at least $1$s away from turn boundaries \cite{ekstedt22_interspeech}. The median FAU intensities visualised in Figure \ref{fig:heatmaps-of} show clear differences in the distribution of visual cues across the two datasets. For instance, speakers in AVCocktail exhibit consistently high intensities in cheek raiser and lid tightener actions across all conditions, including regions where no interlocutor in the dyad is speaking. This may be due to the Lombard effect in visual signals under the effect of noise, as discussed earlier.
\section{Turn-Taking Prediction Experiments}
\label{sec:exps}

\subsection{Model Architectures}
We experiment with various models based on the state-of-the-art stereo Voice Activity Projection (VAP) architecture, first proposed by Ekstedt et al. \cite{ekstedt23_interspeech}. Given a stereo signal $X$ of two speakers $S_1$ and $S_2$ in a dyadic conversation, $X=[X_{S_1}, X_{S_2}]$. At a time step $t$,  the model learns to predict the speaker activity of both speakers in a $2$-second-future window, divided into 8 bins (4 for each speaker) of different sizes, ordered as $200$ms, $400$ms, $600$ms, and $800$ms. The speaker activity label is binary in each bin, totalling $2^8$ possible outputs to represent the activity of both speakers in 8 bins. In a formal definition, the model learns to predict $P(Y_\text{vap}|X) = P(Y_\text{vap}=k| X)$ where $k=1,2,..,K \leq 2^{n \cdot C}$, $n$ and $C$ are the number of bins and number of speakers, respectively. 
To stabilise training, an additional voice activity detection (VAD) task is included to predict the speaker activities $P(Y_{\text{vad}}|X)$ of each interlocutor \cite{inoue2024multilingual}. Note that the VAD task models speaker activity within the present window, whereas the VAP objective predicts activity in a future window beyond the temporal context of $X$.

\begin{figure}
    \centering
    \includegraphics[width=0.68\linewidth]{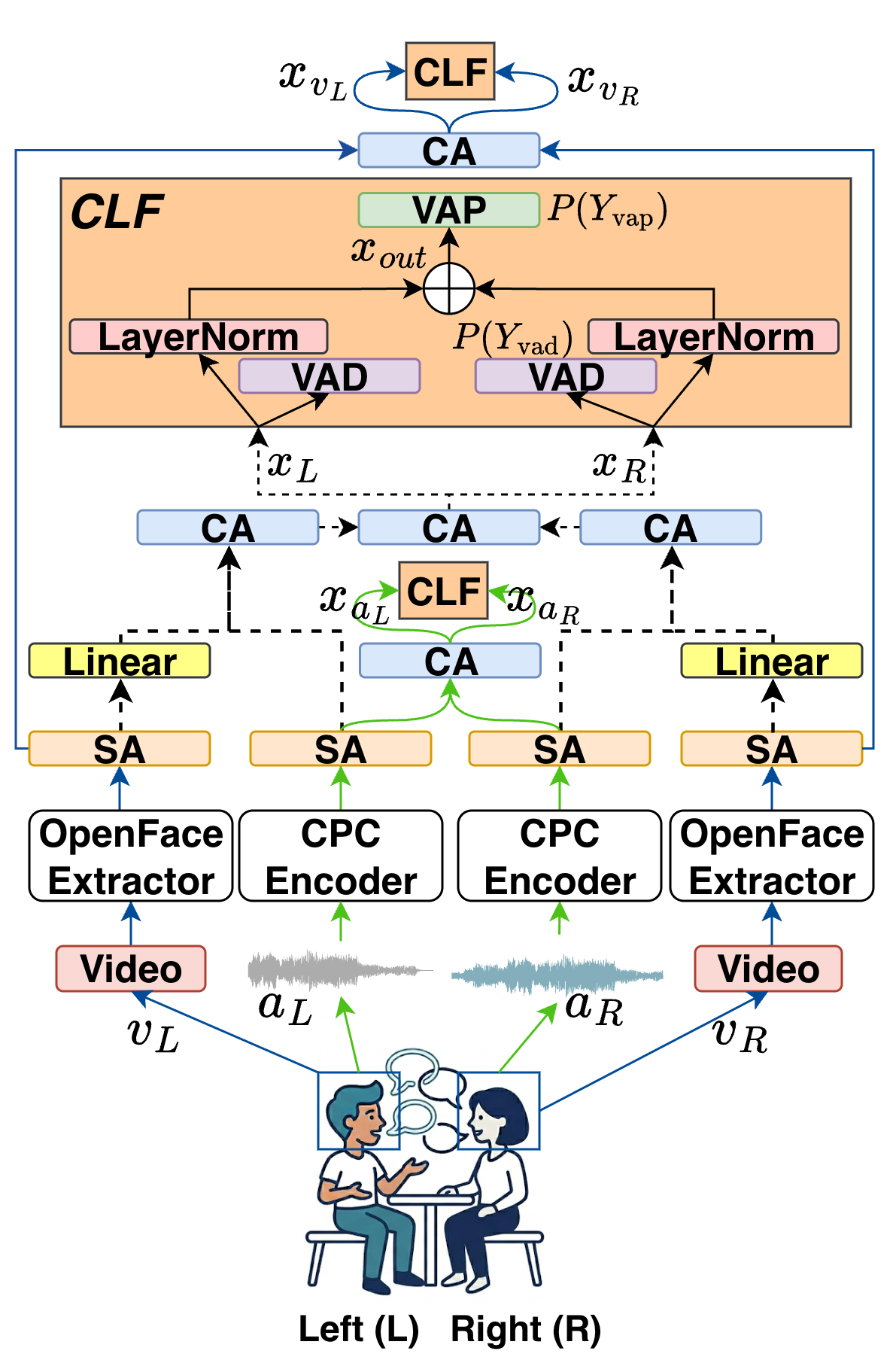}
    \caption{Model architectures of the VAP variants. CA is cross-attention. SA is self-attention.
    Green and blue lines denote the flow of feature extraction and classification for Audio-VAP and Video-VAP, respectively. The MM-VAP model uses audio and video features, with flows in dashed lines for feature fusion. The CLF has the classifier for voice activity projection (VAP) in the future window, and voice activity detection (VAD) for the current context window. The Left (L) and Right (R) notations only represent the view of the model implementations and do not imply the actual setup in the dataset.}
    \label{fig:architecture}
    \vspace{-1em}
\end{figure}

We use three variants of VAP model, with architectures illustrated in Figure \ref{fig:architecture}:
1) \textbf{Audio-VAP:} The original model and VAP architecture \cite{ekstedt23_interspeech}. The input signal $X$ is the stereo audio, where each channel is a separate audio of one interlocutor in the dyadic conversation, i.e. $X \equiv [a_L, a_R]$. The final features for VAD and VAP classification are $[x_L, x_R] \equiv [x_{a_L}, x_{a_R}]$.
2) \textbf{Video-VAP:} The extension of VAP, with video signals as input instead of audio \cite{russell2025visual} 
    , i.e. $X \equiv [v_L, v_R]$.
    The final features for VAD and VAP classification are $[x_L, x_R] \equiv [x_{v_L}, x_{v_R}]$. Both single-modality VAP models use self-attention (SA) \cite{vaswani2017attention} to learn intra-channel representations before being fused into cross-channel information with the cross-attention (CA) \cite{vaswani2017attention}.
3) \textbf{MM-VAP:} The multimodal model \cite{russell2025visual} receives the audio and video signals of both speakers as the input. The SA  is utilised to learn the intra-modality representations, then later fused into an cross-modality embedding with the CA block. Finally, these features from both speakers are fed into a CA block to learn the dependencies between interlocutors in the conversation before the VAD and VAP classifiers. Since the audio and video feature extractors have different embedding dimensions, a linear projection is utilised to project the video features into the same dimension as the audio. 

\subsection{Experiment Setup}
\label{sec:exp-setup}
\subsubsection{Feature Extraction}
We feed sliding windows of $20$ seconds with an overlap of $10$ seconds into the models. All audio is sampled at $16$kHz and normalised. Audio features are extracted at $50$Hz with a pre-trained Contrastive Predictive Coding (CPC) \cite{riviere2020unsupervised} model, as used extensively in previous studies \cite{ekstedt22_interspeech, russell2025visual, ekstedt2022much, inoue2024multilingual}.
Videos are standardised to $50$ FPS with linear interpolation for both datasets to match the frame rate of audio. Video features extracted from OpenFace include eye gaze, head poses, face landmarks, and $17$ action units (detailed in \cite{russell2025visual}). All video features are normalised to the $0-1$ range at the speaker level except for action units intensity, whose range is $1-5$. There are $60$ features extracted in total. In MM-VAP, video features are upsampled to $256$-dimensional embeddings to match the audio dimension for the CA block.

\subsubsection{Dataset Splits}
We use the training and development splits from the first fold of Candor in \cite{russell2025visual, oconnorrussell25_interspeech}.  For AVCocktail, we utilise the provided public splits \cite{nguyen25b_interspeech}. Since the evaluation split of AVCocktail does not feature transcription labels, we cannot deduce the turn labels accordingly and this partition is not used in this study.

\begin{table*}[htbp]
\centering
\begin{threeparttable}

\caption{Performance of VAP models under different training settings. Group I and II show models trained on Candor and AVCocktail, respectively. Group III shows models pre-trained on Candor then fine-tuned on AVCocktail.  \textbf{Bold} denotes the best result per dataset and metric. Numbers in parentheses in Group III show relative changes\tnote{*} compared to the same model in Group I.}
\label{tab:merged-results}
\small
\begin{tabular*}{\textwidth}{@{\extracolsep{\fill}}ccc|rrr|rrr}
\toprule
\textbf{Group} & \textbf{Model} & \textbf{Setting} &
\multicolumn{3}{c|}{\textbf{Candor (F1 \%)}} &
\multicolumn{3}{c}{\textbf{AVCocktail (F1 \%)}} \\

 & & 
 & \textbf{Shift} & \textbf{Hold} & \textbf{Weighted}
 & \textbf{Shift} & \textbf{Hold} & \textbf{Weighted} \\
\midrule

\textbf{-} & Random & - 
& 31.48 & 60.55 & 52.15 
& 45.08 & 45.08 & 45.08 \\

\midrule

\multirow{3}{*}{\textbf{I}} 
& \textbf{MM}-VAP    & Candor 
& 67.07 & \textbf{88.87} & \textbf{82.57}
& 56.24 & 57.05 & 56.61 \\

& \textbf{Audio}-VAP & Candor 
& 65.28 & 87.47 & 81.05 
& 51.48 & 47.84 & 49.83 \\

& \textbf{Video}-VAP & Candor 
& 39.95 & 83.17 & 70.68 
& 67.35 & 71.49 & 69.23 \\

\midrule

\multirow{3}{*}{\textbf{II}} 
& \textbf{MM}-VAP    & AVCocktail 
& 31.18 & 78.14 & 64.56 
& 68.50 & 71.46 & 69.84 \\

& \textbf{Audio}-VAP & AVCocktail 
& 36.90 & 59.66 & 53.08 
& 55.02 & 53.07 & 54.14 \\

& \textbf{Video}-VAP & AVCocktail 
& 29.48 & 77.73 & 63.78 
& 69.38 & 72.14 & 70.63 \\

\midrule

\multirow{3}{*}{\textbf{III}} 
& \textbf{MM}-VAP    & Candor $\rightarrow$ AVCocktail 
& 59.87 & 87.62 & 79.60 ($\downarrow3.60\%$)
& \textbf{72.45} & \textbf{75.10} & \textbf{73.65} ($\uparrow30.10\%$) \\

& \textbf{Audio}-VAP & Candor $\rightarrow$ AVCocktail 
& 61.99 & 86.95 & 79.74 ($\downarrow 1.62\%$)
& 50.22 & 44.19 & 47.49 ($\downarrow4.70\%$) \\

& \textbf{Video}-VAP & Candor $\rightarrow$ AVCocktail 
& 28.72 & 82.19 & 66.74 ($\downarrow5.57\%$)
& 71.81 & 74.46 & 73.01 ($\uparrow5.46\%$) \\

\midrule
\multirow{4}{*}{\textbf{IV}}
& \textbf{MM}-VAP & Candor$^\dagger$  & \textbf{67.95} & 88.47 & 82.54 & 58.55 & 62.97 & 60.55 \\
& \textbf{MM}-VAP & Candor$^\dagger$ $\rightarrow$ AVCocktail   & 57.85 & 87.56 & 78.97 & 69.06 & 72.35 & 70.55 \\

& \textbf{Audio}-VAP & Candor$^\dagger$  & 61.43 & 87.50 & 79.97 & 52.11 & 47.22 & 49.89 \\
& \textbf{Audio}-VAP & Candor$^\dagger$ $\rightarrow$ AVCocktail   & 57.84 & 87.09 & 78.64 & 51.99 & 46.32 & 49.42 \\


\bottomrule
\end{tabular*}
\begin{tablenotes}
\item[*]All changes are significant ($p<0.05$) in t-tests with recording sessions as the unit of analysis.  $^\dagger$Candor with 25\% of sessions augmented with noise (music, babble, speech) at $0dB$ SNR \cite{oconnorrussell25_interspeech}.
\end{tablenotes}
\end{threeparttable}
\vspace{-2em}
\end{table*}

\subsubsection{Training Configurations}
The Audio-VAP, Video-VAP, and MM-VAP checkpoints trained on the full $601.81$h of the Candor dataset are taken from the open-source repository of the study by Russell and Harte \cite{russell2025visual}. All other experiments are run on a single NVIDIA RTX A4000, conducted for 10 epochs with a batch size of 4, and the epoch with the best performance on the development set is reported. For fine-tuning, we schedule the learning rate to increase linearly from $0$ to $3\times10^{-5}$ in the first $2,000$ training steps and remain constant after that, to stabilise the initial phase of the adaptation. 
Training on AVCocktail uses a constant learning rate of $3\times10^{-5}$.
All training is optimised with the AdamW \cite{loshchilov2018decoupled} algorithm. The VAP classifier is optimised with the cross-entropy loss function, the VAD classifier is optimised with binary cross-entropy loss. The final optimisation loss is the sum of VAD and VAP losses.
In the SA module, we use 4 attention heads and one attention layer, with an embedding dimension of $256$. For CA blocks, we use $4$ heads and $3$ attention layers, with the embedding dimension of $256$.

\subsubsection{Evaluation Configurations}
We follow the configurations in \cite{russell2025visual}. All models make predictions at $50$Hz. At the starting timestamp of every shift or hold in the validation set, we compute the cumulative shift probability in the next $200$ms. If the probability exceeds a threshold, a shift event is predicted, else hold. Without loss of generality, we choose the cumulative threshold as $0.5 \times W \times f$, where $f$ is the prediction rate ($50$ Hz), and $W$ is the prediction window size ($0.2$ seconds). We report the F1 score for shift and hold events, and weighted F1.
\section{Results and Discussion}
\label{sec:results}

\subsection{Generalisation of Turn-taking Models}

Group I of Table \ref{tab:merged-results} details results of Candor-trained models, evaluated on Candor and AVCocktail. 
The random baseline provides a useful lower bound. 
Models trained on Candor suffer a pronounced degradation in highly overlapped speech and noisy acoustic conditions like AVCocktail. The multimodal MM-VAP drops by roughly one-third in weighted F1, from $82.57\%$ to $56.61\%$, with a particularly sharp decline in turn-holding cues detection ($88.87\%$ to $57.05\%$). This phenomenon may be attributed to the fact that during short pauses, other parallel conversations in the session still progress, which confuses the models if they over-rely on silences to detect event cues. 
Audio-VAP is the most affected: weighted F1 ($81.05\%$ on Candor) and hold F1 ($87.47\%$) both decrease to only above-random when evaluated on AVCocktail, suggesting strong sensitivity to acoustic mismatch and noise. 
In contrast, Video-VAP remains comparatively stable across domains, with a weighted F1 of $70.68\%$ and $69.23\%$ on Candor and AVCocktail respectively ($p=0.3689$ in independent t-test). However, its overall performance remains limited, lagging behind Audio-VAP and MM-VAP for in-domain Candor evaluation. This trend indicates that using just visual cues is insufficient for high-quality turn-taking prediction. Besides, there is a need for more contextualised, lower-level features rather than the high-level OpenFace features to push the performance of visual-based turn-taking predictors in future research.
















For models trained on AVCocktail, shown in Group II of Table \ref{tab:merged-results}, the trends reflect both the data scale and the task complexity. Given that AVCocktail is about $100$ times smaller than Candor and contains highly complex acoustic conditions, Audio-VAP does not learn sufficiently robust representations. This yields a relatively poor performance of only $54.14\%$ weighted F1 even in the same domain. Video-VAP trained on AVCocktail achieves comparable results across both datasets ($63.78\%$ for Candor and $70.63\%$ for AVCocktail), although its performance is less robust compared to that trained on Candor with a relative drop of $8.75\%$ in weighted F1 on average. This is possibly due to the reduced training data. The multimodal MM-VAP strikes the best balance, achieving comparable performance across both Candor ($64.56\%$) and AVCocktail ($69.84\%$). Overall, these results reinforce that audio is highly domain-sensitive, video is more stable but limited, and multimodality provides a more complete solution.

\subsection{Audio-Visual Adaptation of Turn-taking}












Group III of Table \ref{tab:merged-results} details the performance of VAP variants when pre-trained on Candor, and fine-tuned on AVCocktail. Fine-tuning on the target domain consistently boosts performance. MM-VAP improves its relative weighted F1 on AVCocktail by roughly $30\%$ after fine-tuning, and remains better than the AVCocktail-trained model ($73.65\%$ and $69.84\%$). The Video-VAP model also observes a similar trend, improving from $69.23\%$ to $73.01\%$ in weighted F1.
Meanwhile, the audio-only variant fails to adapt to the new domain effectively, with a relative drop in weighted F1 of $4.70\%$, from $49.83\%$ to only above-random. 
These trends jointly show that video offers a robust signal for predicting turn-taking cues, and when visual cues are learned in tandem with the relevant audio features, the performance improves even further.

It is worth noting that the results on AVCocktail when MM-VAP and Video-VAP are trained or fine-tuned on AVCocktail (Group II and III in Table \ref{tab:merged-results}) are not significantly different ($p>0.05$). This observation suggests that, under complex acoustic conditions, the audio modality provides limited information compared to visual cues. One possible explanation is that the current multimodal fusion mechanism is unable to effectively exploit cross-modal dependencies when the audio signals are highly overlapped or noisy, causing the model to rely more heavily on visual information. These findings highlight the need for more robust fusion strategies in future multimodal turn-taking prediction systems.

Improvements after fine-tuning are accompanied by catastrophic forgetting, as seen with Candor after fine-tuning in Group III of Table \ref{tab:merged-results}. After fine-tuning on AVCocktail,  Candor performance consistently drops by $3–5\%$ relative to the pre-finetuning models, showing a clear trade-off between specialising in the new domain and retaining old knowledge. Visual-supported models balance this trade-off more effectively, given that the relative loss of MM-VAP and Video-VAP on Candor after fine-tuning is small (under $6\%$), while their relative gains on AVCocktail remain larger (up to $30.10\%$ in MM-VAP and $5.46\%$ in Video-VAP). Thus, visual cues improve generalisation and make the model less susceptible to forgetting.

\begin{figure}
    \centering
    \includegraphics[width=0.9\linewidth]{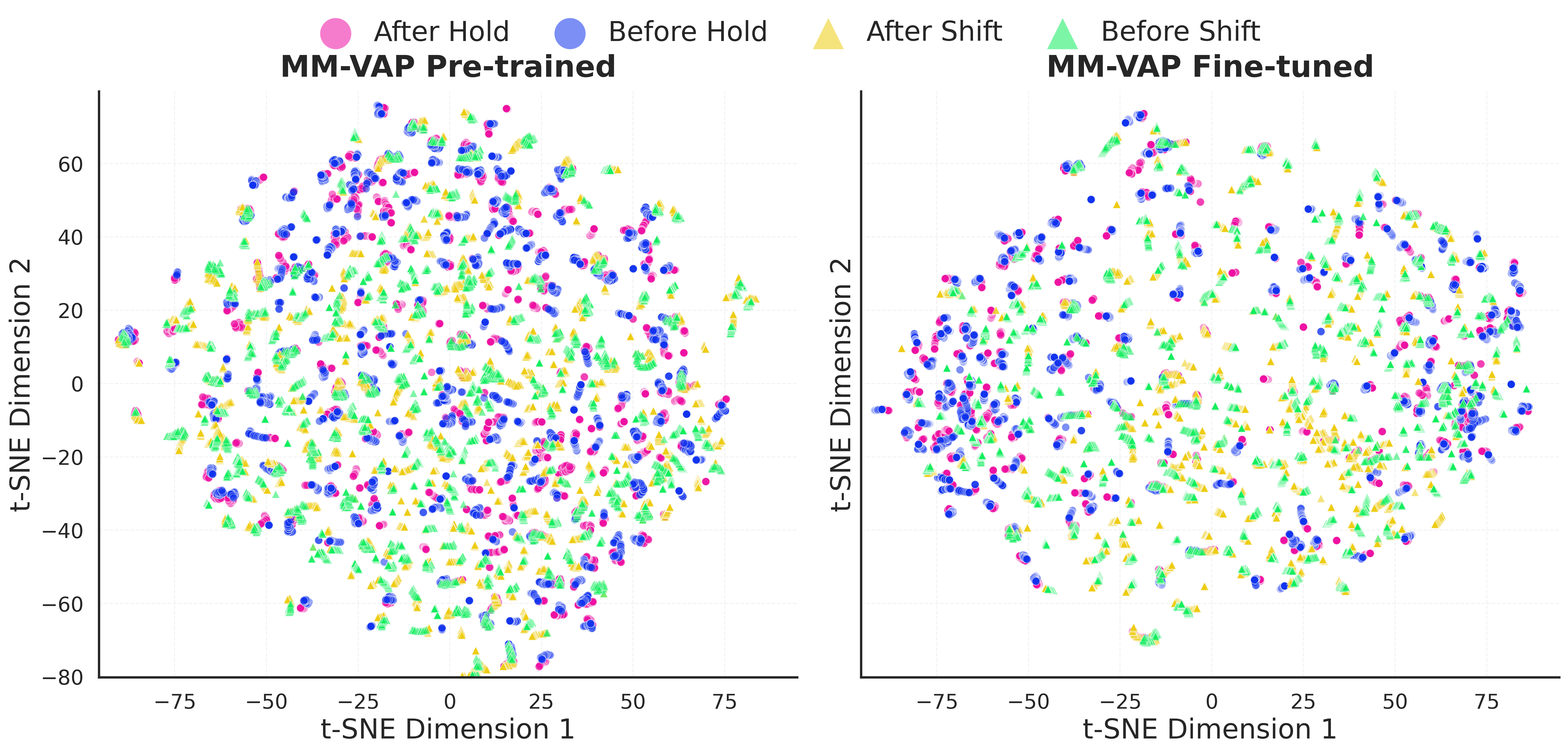}
    \caption{t-SNE visualisations of embeddings from MM-VAP, $1$s before and after every shift and hold. Segments from $30^\text{th}$-$60^\text{th}$ sec in AVCocktail-dev.}
    
    \label{fig:embeddings}
    \vspace{-1em}
\end{figure}

To examine how fine-tuning affects the learned representation, we extract embeddings before the VAP classifier in MM-VAP ($x_{out}$ in Figure~\ref{fig:architecture}). Using the AVCocktail-dev set, embeddings are collected one second before and after each shift and hold event from the $30^\text{th}$–$60^\text{th}$ seconds of all recordings.
They are then visualised using t-SNE \cite{van2008visualizing} with $3{,}000$ optimisation iterations and a perplexity of $150$, chosen to emphasise separation between shift and hold events across recordings. Figure~\ref{fig:embeddings} compares embeddings from MM-VAP pre-trained only on Candor (left) and the same model further fine-tuned on AVCocktail (right).
The Candor-only model fails to clearly separate shift and hold events in AVCocktail, with embeddings remaining heavily intertwined. After fine-tuning, the two event types become more distinguishable, and embeddings before and after the same event are drawn closer together, suggesting improved temporal consistency. Although the separation remains imperfect due to the challenging cocktail-party setting, the fine-tuned model shows a clearer organisation of turn-taking cues.

 We explore the approach of training with actual noise rather than augmenting training with synthetic noise, as studied in \cite{oconnorrussell25_interspeech, inoue2025noise}. We use the Audio-VAP and MM-VAP models trained on Candor augmented with $0dB$ synthetic noise \cite{oconnorrussell25_interspeech}, which can be babble, music, or speech noise. As seen in Group IV-Table \ref{tab:merged-results}, the addition of synthetic noise improves the results on AVCocktail compared to training on clean Candor only, while not harming the performance on the original domain. However, using synthetic noise lags behind training with the actual noisy data of AVCocktail: $60.55\%$ vs. $69.84\%$ for MM-VAP, and $49.89\%$ vs. $54.14\%$ for Audio-VAP. Fine-tuning on AVCocktail from a pre-training with augmented noise does not benefit the adaptation since the changes in AVCocktail compared to Group III are marginal.



\subsection{Effect of Pre-training Dataset Size}
\label{sec:pretrain-scale}
\begin{figure}[htbp]
\vspace{-1em}
    \centering
    \includegraphics[width=0.8\linewidth]{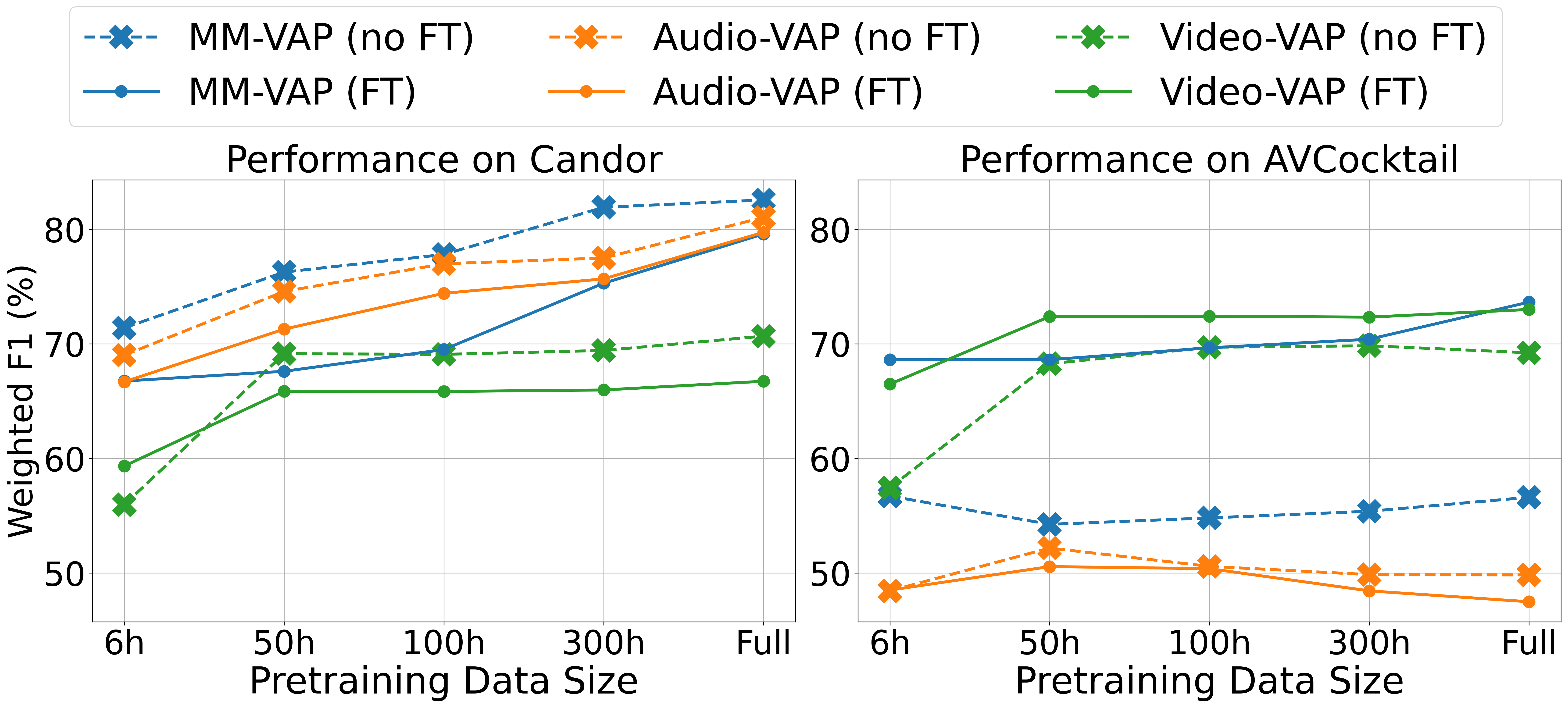}
    \caption{Performance before (dashed lines) and after fine-tuning (solid lines), with increasing pre-training data.  AVCocktail has 6h, full Candor is 600h.}
    \label{fig:scale-plots}
    \vspace{-1em}
\end{figure}
We have shown that to generalise to a new domain, turn-taking models benefit from large-scale pre-training to learn informative audio-visual turn-taking cues, and benefit from subsequent fine-tuning on a smaller in-domain dataset.
To investigate how much pre-training data is required for effective adaptation, we sample the original Candor dataset of $600$h into smaller subsets for pre-training: $6$h (roughly AVCocktail size), $50$h, $100$h, and $300$h (half the original size). The duration distribution of the recording sessions is first discretised into 10 bins of duration range. The subsets are sampled so that the duration distribution matches the original. These pre-trained models (no FT) are all then fine-tuned on AVCocktail (FT). 
Figure \ref{fig:scale-plots} shows how  performance on Candor (left) and AVCocktail (right) is impacted, as the pre-training scale is varied. 
The dashed lines (no FT) in the left plot show a clear benefit from scaling up the Candor training data.
From $6$h to the full $600$h, MM-VAP and Audio-VAP model gain around $10\%$ in absolute weighted F1 on Candor.
Video-VAP peaks at $70\%$ with $50$h of training data and then plateaus. It consistently lags behind Audio-VAP and MM-VAP. 




The solid lines in Figure \ref{fig:scale-plots} (right) show the results on AVCocktail after fine-tuning. Increasing the pre-training size also improves the downstream fine-tuning performance, and the effect differs across modalities. The fine-tuned Video-VAP shows the largest gap between $6$h ($67\%$ weighted F1) and the full $600$h ($73\%$), again with the steepest change at $50$h of pre-training. The fine-tuned MM-VAP improves more steadily with the pre-training size, starting at nearly $70\%$ weighted F1 with $6$h of pre-training data, and gradually increases to roughly $74\%$. 
In contrast, Audio-VAP struggles with the adaptation. Performance on AVCocktail does not improve, and even decreases, with different pre-training sizes. 




The drop in performance on Candor after fine-tuning on AVCocktail (see solid lines in Figure \ref{fig:scale-plots}-left)  shows that catastrophic forgetting is a consistent issue. The effect is smaller in Audio-VAP and consistently reduces with more pre-training data. However, its performance on AVCocktail drops close to random, with or without fine-tuning. This suggests the audio model is not adapting its internal parameters when the new domain is too noisy. For MM-VAP, this forgetting phenomenon consistently reduces as the pre-training scales up, showing effective adaptation to the cocktail-party scenario.





\section{Conclusions}
\label{sec:conclusion}

To be useful, turn-taking prediction systems must generalise to realistic, noisy conversational settings. In this work, we present the first evaluation of turn-taking models in a cocktail-party scenario, using the AVCocktail dataset. Our results show that pre-trained VAP models, across different modalities, are highly sensitive to domain shift, with performance degrading by $2\%$ to $38\%$ relatively in weighted F1 under background noise and complex interaction dynamics. When VAP models are pre-trained on a sufficiently large dataset to capture the basic dynamics of turn-taking, fine-tuning can improve the robustness in the new domain, especially with the rich contextualised information from both audio and visual cues, with up to $30\%$ relative improvement. However, adaptation also introduces catastrophic forgetting of the original domain, with relative drops from $1.6\%$ to $5.6\%$ across modalities. 
Experimental results with multimodal models indicate that audio-visual cues are essential not only for mitigating this forgetting, but also for more effective exploitation of turn-taking signals across different pre-training data sizes. Results also reveal limitations in the current approaches to turn-taking: high-level features like OpenFace are not optimal for high-quality turn-taking systems; and the current architecture to fuse audio and visual information does not exploit the complementarity between modalities effectively.

\section*{Acknowledgement}

This paper emanates from research supported by Taighde Éireann – Research Ireland, Grant number 22/FFP-A/11059.

\section{Generative AI Use Disclosure}
During the preparation of this work, ChatGPT (OpenAI) was used only for minor English grammar corrections and refining the clarity of written content.
\bibliographystyle{IEEEtran}
\bibliography{turn-taking}
\newpage
\appendix[Manual Turn Labels Verification of AVCocktail-dev]

\begin{figure}[h]
    \centering
    \includegraphics[width=\linewidth]{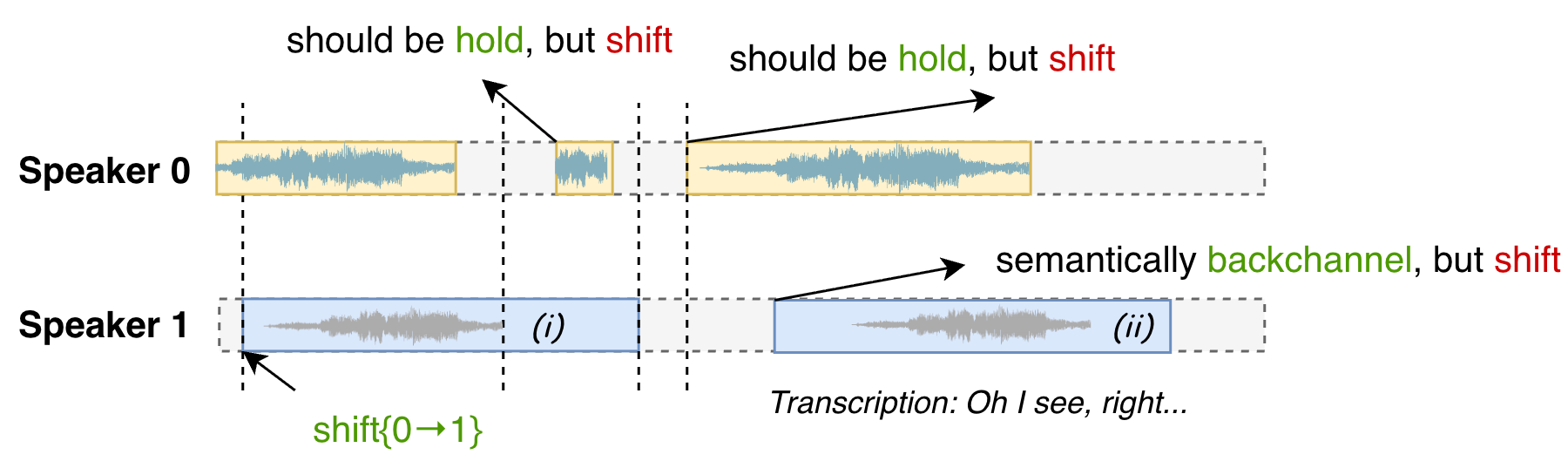}
    \caption{Examples of failure cases identified after manual verification. Events in red are automatic labels (Section \ref{sec:avcocktail-dataset}), and green shows the corrected events. (i) shows wrong timestamps for speech endpoints of \texttt{speaker 0}, leading to a wrong decision of events in \texttt{speaker 1}; (ii) shows both wrong timestamp endpoints of the backchannel and the limitation of rule-based decision when it comes to long IPUs that are semantically backchannels and show no intention of taking the conversation floor.}
    \label{fig:avcocktail-fails}
\end{figure}
\begin{table}[h]
\small
\centering
\caption{Statistics of the AVCocktail-dev subset after manual verification.}
\label{tab:avcocktail-dev-annotated}
\begin{tabular}{lrrr}
\toprule
\textbf{Subset} & \textbf{\# Dyads} & \textbf{\# Shifts} & \textbf{\# Holds} \\ \midrule
AVCocktail-dev  & 33                & 1,916              & 1,758             \\ \bottomrule
\end{tabular}
\end{table}

\begin{table*}[htbp]
\centering
\begin{threeparttable}

\caption{Performance of VAP models on AVCocktail-dev with manually verified labels. Group I and II show models trained on Candor and AVCocktail, respectively. Group III shows models pre-trained on Candor then fine-tuned on AVCocktail. \textbf{Bold} denotes the best result per metric. Numbers in parentheses show absolute changes compared to the automatic labels (Table~\ref{tab:merged-results}).}
\label{tab:annotated-results}
\small
\begin{tabular*}{\textwidth}{@{\extracolsep{\fill}}ccc|rrr}
\toprule
\textbf{Group} & \textbf{Model} & \textbf{Setting} &
\multicolumn{3}{c}{\textbf{AVCocktail-dev (F1 \%)}} \\

 & &
 & \textbf{Shift} & \textbf{Hold} & \textbf{Weighted} \\
\midrule

\multirow{3}{*}{\textbf{I}}
& \textbf{MM}-VAP    & Candor
& 53.61 ($\downarrow2.63$) & 57.49 ($\uparrow0.44$) & 55.47 ($\downarrow1.14$) \\

& \textbf{Audio}-VAP & Candor
& 50.85 ($\downarrow0.63$) & 49.33 ($\uparrow1.49$) & 50.12 ($\uparrow0.29$) \\

& \textbf{Video}-VAP & Candor
& 68.47 ($\uparrow1.12$) & 75.02 ($\uparrow3.53$) & 71.60 ($\uparrow2.37$) \\

\midrule

\multirow{3}{*}{\textbf{II}}
& \textbf{MM}-VAP    & AVCocktail
& 73.62 ($\uparrow5.12$) & 77.08 ($\uparrow5.62$) & 75.28 ($\uparrow5.44$) \\

& \textbf{Audio}-VAP & AVCocktail
& 51.42 ($\downarrow3.60$) & 50.59 ($\downarrow2.48$) & 51.02 ($\downarrow3.12$) \\

& \textbf{Video}-VAP & AVCocktail
& 72.62 ($\uparrow3.24$) & 76.71 ($\uparrow4.57$) & 74.58 ($\uparrow3.95$) \\

\midrule

\multirow{3}{*}{\textbf{III}}
& \textbf{MM}-VAP    & Candor $\rightarrow$ AVCocktail
& \textbf{76.77} ($\uparrow4.32$) & \textbf{80.25} ($\uparrow5.15$) & \textbf{78.43} ($\uparrow4.78$) \\

& \textbf{Audio}-VAP & Candor $\rightarrow$ AVCocktail
& 50.01 ($\downarrow0.21$) & 45.95 ($\uparrow1.76$) & 48.07 ($\uparrow0.58$) \\

& \textbf{Video}-VAP & Candor $\rightarrow$ AVCocktail
& 74.49 ($\uparrow2.68$) & 78.45 ($\uparrow3.99$) & 76.38 ($\uparrow3.37$) \\

\bottomrule
\end{tabular*}

\end{threeparttable}
\vspace{-2em}
\end{table*}

\begin{figure}[h]
    \centering
    \includegraphics[width=\linewidth]{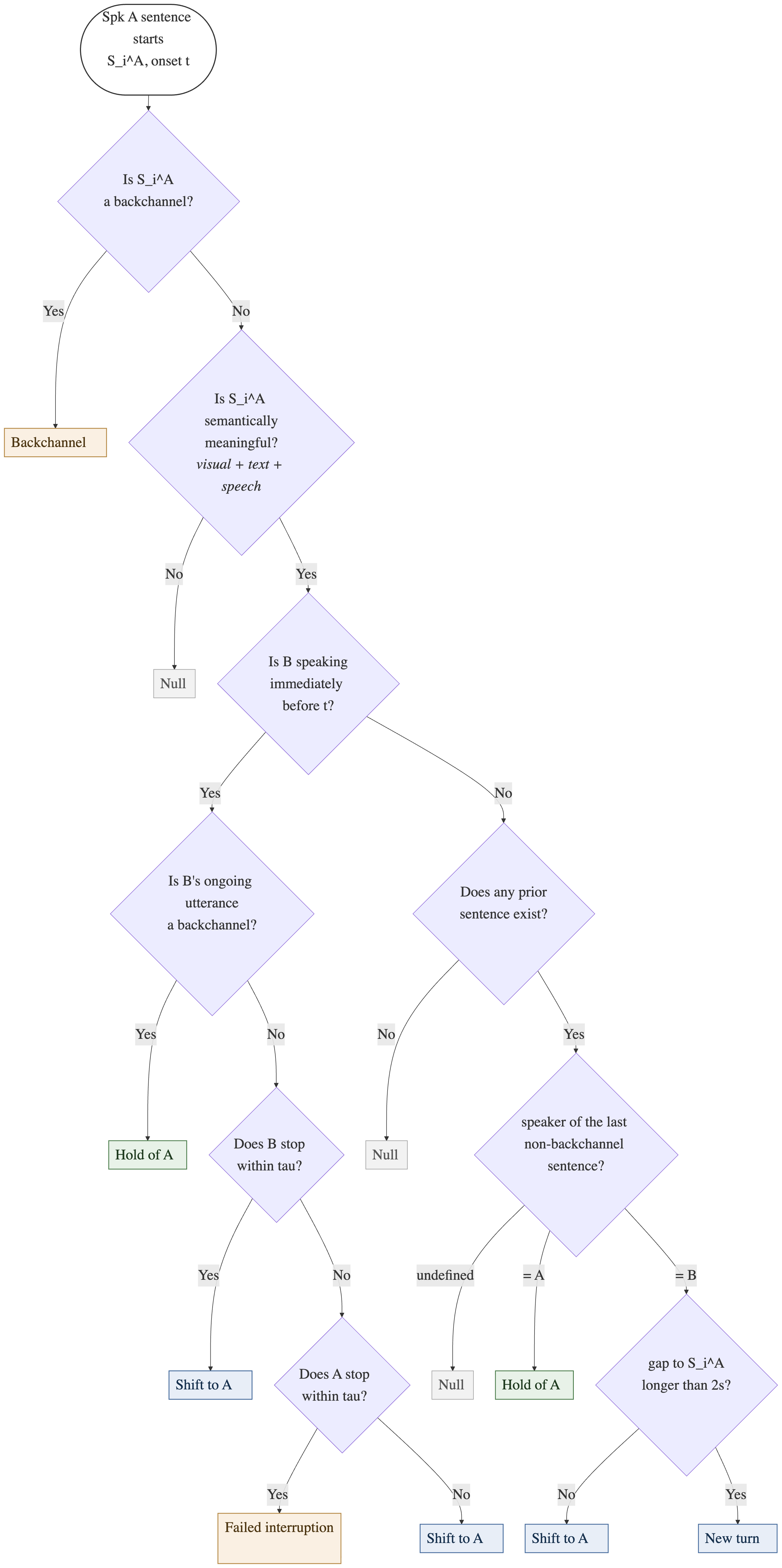}
    \caption{Diagram showing the decision process of shift/hold label of a sentence $S_i^A$ of speaker A. Semantically meaningfulness are decided subjectively by the annotator, based on whether the speaker is showing (speech and visual) cues to take or maintain the floor or not. Failed interruptions are also labelled to facilitate further research. Null means there is no appropriate event labels at that timestamp. A new turn is neither a shift nor a hold, and is established after a long ($>\tau$) mutual silence without intention of maintaining the floor from any speaker.}
    \label{fig:diagram}
\end{figure}
We manually verify automatic turn labels in AVCocktail-dev, generated by the definition in Section \ref{sec:avcocktail-dataset}. The annotation is done by the authors with the ELAN \cite{sloetjes2008annotation} framework. At each event, we verify and correct to shift/hold, adjust timestamps if needed, and annotate failed interruptions and backchannels to facilitate further research. The labels are decided based on the decision diagram illustrated in Figure \ref{fig:diagram}. The statistics of shift and hold are described in Table \ref{tab:avcocktail-dev-annotated}. We remove two sessions which are not dyadic conversations after manual inspection. It can be seen that the number of holds does not change much (1,758 after manual check compared to 1,757). Variances in shifts can be attributed to various factors with examples shown in Figure \ref{fig:avcocktail-fails}. The most common scenario is backchannels mistakenly decided as shifts. Backchannels ("uh", "hmm", etc.) may have annotated timestamps spread over 1 second despite being just short vocalisations, which are not effectively ignored by our definition in \ref{sec:avcocktail-dataset}. Backchannels can also come in the form of short answers ("I see", "I agree with you", etc.) with no intention of taking the conversation floor. Without an external semantic understanding module, these errors are ignored, while manual verification can handle them. Another common factor is incorrect timestamps. Figure \ref{fig:avcocktail-fails} (i) shows an example where speaker 1's ending timestamp is too late compared to the actual speech offset, leading to a false shift for speaker 0 and propagating further.

The turn-taking prediction results on the annotated AVCocktail are shown in Table \ref{tab:annotated-results}. The relative ranking remains the same compared to automatic labels. We also see a larger gap between MM-VAP and Video-VAP performance, strengthening the advantages of multimodal learning for turn-taking prediction in complex scenarios.

\end{document}